\documentclass[]{emulateapj}

\shorttitle{History of Radio Activity in the Core of A2597}
\shortauthors{Clarke et al.}

\begin{document}

\def\msun{\mbox{$M_\odot$}}

\title{Low Frequency Radio Observations of X-ray Ghost Bubbles in Abell
2597: A History of Radio Activity in the Core}

\author{T.\ E.\ Clarke\altaffilmark{1,2}, C.\ L.\ Sarazin\altaffilmark{1}, E.\ L.\ Blanton\altaffilmark{3}, D.\ M.\ Neumann\altaffilmark{4}, and N.\ E.\ Kassim\altaffilmark{2}}
\altaffiltext{1}{Department of Astronomy, University of Virginia, P. O. Box 3818, Charlottesville, VA 22903-0818, USA}
\altaffiltext{2}{Naval Research Laboratory, Code 7213, 4555 Overlook Ave. SW, Washington, DC  20375, USA}
\altaffiltext{3}{Astronomy Department, Boston University, 725 Commonwealth Ave., Boston, MA 02215, USA}
\altaffiltext{4}{CEA/Saclay, Service d'Astrophysique, L'Orme des Merisiers, B\^at.\ 709, 91191 Gif-sur-Yvette Cedex, France}

\begin{abstract}
A previous analysis of the {\it Chandra} X-ray image of the center of
the cooling core cluster Abell 2597 showed two ``ghost holes'' in the
X-ray emission to the west and northeast of the central radio galaxy
PKS~2322$-$123.
Previous radio observations did not detect any radio
emission coming from the interior of the X-ray holes.  We present new
low frequency radio observations of Abell~2597.  At 330 MHz, radio
emission extends into the interior of the western ghost bubble, but
not the northeast one.  Our re-analysis of the archival {\it Chandra}
data shows evidence for an X-ray tunnel (elongated region of reduced
X-ray emission) extending from near the center of the cD out to the
west ghost bubble.  We also detect a smaller X-ray hole to the
northeast of the center of the cD and closer than the outer ghost
bubbles.  Radio observations at 1.3 GHz show extensions to the west
along the X-ray tunnel toward the west ghost bubble, to the northeast
into the new X-ray hole, and to the northwest.  All of these
structures are much larger than the two inner radio lobes seen
previously at 8 GHz.  The X-ray tunnel suggests that the west ghost
bubble is part of a continuous flow of radio plasma out from the active galactic nucleus,
rather than a detached buoyant old radio lobe, and 
thus it may be an intermediate case between an active radio galaxy and
a buoyant lobe.
\end{abstract}


\section{Introduction}

The thermal gas in galaxy clusters provides a historical record of the
activity within the last $\sim 10^8$ yr in the cluster. Without recent
(significant) merger activity, the cluster gas relaxes to hydrostatic
equilibrium. In a relaxed cluster, the thermal gas distribution
is relatively smooth and symmetric in the outer cluster regions, while
the inner regions show very peaked X-ray emission which has
often been interpreted as a cooling-flow \citep{af94}.  The inner
regions of these clusters generally contain large cD galaxies, which
usually show evidence for cooler gas (e.g., optical line emission) and
star formation.  These central cD galaxies typically host powerful
radio sources which, in some cases, appear to trigger star
formation. Recent $Chandra$ observations have revealed that there also
appears to be a complex interplay between the central radio sources
and the thermal intracluster medium (ICM). 

X-ray observations of clusters such as Perseus
\citep{bohringer93,perseus}, Hydra A \citep{mcnamara00}, A2052
\citep{blanton01}, and Centaurus \citep{taylor02} with ROSAT and
$Chandra$ show depressions (or bubbles) in the thermal gas that appear
to be spatially coincident with the radio lobes.
The interplay between the thermal and radio
plasma is complex. Contrary to predictions of supersonic expansion of
radio sources \citep{heinz98}, 
the X-ray observations of these clusters do not reveal the
presence of strong shocks near the radio lobes. The radio sources
appear to expand subsonically or mildly transonically into the ICM and
slowly displace the thermal gas \citep{perseus,mcnamara00}.  This
displacement of the X-ray gas is thought to lead to the bright rims of
cool gas observed along the edges of the radio lobes. 
The cool nature
of this gas has been determined by the observed soft X-ray spectrum
(Perseus, \citealt{perseus}; A2052, \citealt{blanton03}).  
In turn, the dense cluster medium is thought to confine the radio
source and produce the compact, distorted morphology typical of
cooling core systems.  Equipartition arguments applied to these
systems suggest that the minimum energy pressure in the radio lobes is
an order of magnitude less than the surrounding thermal gas pressure
(Hydra A, \citealt{mcnamara00}; Perseus, \citealt{perseus}; A2052,
\citealt{blanton01}; \citealt{2004ApJ...607..800B}).  Without some
form of internal pressure support (such as hot, diffuse thermal gas)
these X-ray depressions would collapse on sound crossing timescales of
$\sim 10^7$ yr.

In a few clusters, X-ray observations have revealed the presence of
cavities (``ghost holes'') in the thermal ICM which are located well
beyond the radius of the currently active central radio galaxy (e.g.,
Perseus, \citealt{fabian02}; A2597, \citealt{mcnamara01}; A4059,
\citealt{heinz02}).  These ghost holes are thought to be buoyant lobes
from a past outburst of the radio galaxy which have displaced the
thermal gas as they rose through the cluster atmosphere. Observational
evidence suggests that radio galaxies undergo episodic outbursts which
last $\sim 10^7$ yr and have repetition intervals of $\sim 10^8$ yr
\citep[e.g.,][]{mcnamara01}.  
Over the cluster lifetime, the activity
of the central radio galaxy would produce many generations of radio
lobes. 
Each generation of buoyant bubbles may entrain cool cluster gas as
it moves outward in the cluster atmosphere, thus lowering the total
mass inflow in the cluster core. In addition, recent observations of
Perseus \citep{fabian03} and M87 \citep{forman04} show ripples that
suggest that each generation of radio outburst sends weak shocks
through the ICM. These shocks dissipate energy which may be sufficient
to balance the effects of radiative cooling in the cluster cores. In
this scenario, episodic central radio sources may provide the solution
to the ``cooling flow problem'' (the lack of sufficient quantities of
cool gas in the cores of dense clusters). 
Due to a balance between buoyancy and drag forces, the bubbles will rise at a
fraction of the sound speed in the thermal gas; thus, the locations of
the bubbles can be used to place constraints on the lifecycle of the
central radio galaxy. 
Buoyant bubbles also provide a means of
transporting magnetic fields from the central active galactic nucleus to the outer cluster
regions where the fields have been detected to radii of over 500 kpc
\citep{feretti, clarke01}.  
The bubbles also contain aging
relativistic particles which may be re-accelerated during a cluster
merger to produce the diffuse radio relics which have been observed in
some systems \citep{egk01}.

In this paper we present results of a radio and X-ray analysis of the
cluster Abell~2597. This cluster is a nearby, $z=0.0852$, richness
class 0 system which contains a compact ($\sim$ 5\arcsec) central
radio source (PKS~2322$-$123) at 8 GHz \citep{sarazin95}.  Recent {\it
Hubble Space Telescope (HST)} FUV observations of the system reveal
complex morphology with both diffuse emission as well as bright knots
and filaments which appear to trace the 8 GHz radio emission
\citep{odea04}.  $Chandra$ X-ray observations of the system show
irregular X-ray morphology with a bright central region and two low
surface brightness ``ghost'' cavities 18\arcsec\ to the west and
16\arcsec\ to the northeast of the core \citep{mcnamara01}.

Throughout this paper we adopt WMAP cosmological parameters
\citep{wmap} $H_0$ = 71 km s$^{-1}$ Mpc$^{-1}$,
$\Omega_{\Lambda}=0.73$, and $\Omega_m=0.27$.  At the redshift of
Abell~2597, this corresponds to a scale of 1.54 kpc/arcsec.  

\section{Observations and Data Reductions}
\label{obs_red}

\subsection{X-ray}

The central region of Abell 2597 was observed with {\it Chandra} on
2000 July 28 for a total of 39.4 ksec. The observations were taken in
Faint (F) mode with the cluster center on the back-illuminated S3
chip. The ACIS-236789 chips were on and operating at a temperature of
$-120$ C and a frame time of 3.2 s.  The archival observation (OBSID
922) was extracted and reprocessed using CIAO v3.0.2, and CALDB v2.26.
Only events with $ASCA$ grades 0, 2, 3, 4, and 6 were included in the
analysis. The observations did not include data from the other
back-illuminated (S1) CCD, hence the flare filtering was undertaken
using the S3 chip.  Regions surrounding the bright part of the cluster
and compact sources were excluded when determining the light curve.
As discussed previously by \citet{mcnamara01}, the observations were
severely impacted by flares. In an effort to extract as much
information as possible from the data, we have not followed the
standard flare-filtering routines.  In this paper, we concentrate on
the $Chandra$ images of the bright central region of the cluster, and
do not undertake spectral fits; thus, we have not filtered out all
data impacted by flares.  We have only clipped off the strongest parts
of the flares but still include data with up to 8.5 times the mean
quiescent background rate.  This filtering retains $\sim$ 32 ksec of
data, which gives enough photons to examine the detailed spatial
structure of the thermal gas in the central regions.

We examined the impact of the high cutoff for the flare filtering by
determining the contribution of the background counts to the counts in
the central region of the cluster discussed in this paper. Using the
standard flare-filtering cutoff we estimated the total counts in the
off-source region of the S3 chip. Due to the large size of clusters,
this region will contain both cluster and background counts. We used
the quiescent\footnote{Quiescent background estimate for S3 taken from
ACIS background memo at http://cxc.harvard.edu/contrib/maxim/bg.}
background estimate of 0.86 cts/s/chip for the S3 chip (properly
scaled to the relevant area) to subtract the quiescent background and
determine the cluster contribution in the $0.3-10$ keV band. This
contribution was removed from the counts determined for the off-source
region in our relaxed flare-filtering data set. Scaling the remaining
quiescent plus background flare counts to the area of interest in this
paper, the background is found to contribute only 1.1\% of the source
counts. Given this very small contribution to the counts and the
overall spatial uniformity of the background flares, the spatial
structure of the features discussed below are not influenced by our
relaxed filtering criteria.

\subsection{Radio}

\begin{deluxetable}{lcccc}
\tablecaption{Radio Observations of PKS~2322$-$123}
\tablewidth{0pt}
\tablehead{
\colhead{Date} & \colhead{Array}   & \colhead{Frequency}   &
\colhead{Bandwidth} &
\colhead{Duration}\\
\colhead{} & \colhead{} & \colhead{(MHz)} & \colhead{(MHz)} & \colhead{(hours)}
}
\startdata
1996 Dec 7  & A & 1312.15 & 25.0 & 5.0\\
1996 Dec 7  & A & 4985.1 & 50.0 & 1.3\\
2002 Jun 10 & B & 321.5/328.5 & 6.25/6.25 & 2.3 \\
2003 Aug 18 & A & 328.5 & 6.25 & 3.0 \\
\enddata
\label{tbl:radio_obs}
\end{deluxetable}

We observed the central radio galaxy PKS~2322$-$123 in Abell~2597 at
330 MHz with the Very Large Array (VLA).  The observations were taken
in spectral line mode to reduce the effects of bandwidth smearing and
enable radio-frequency interference (RFI) excision.  The observations
around 330 MHz had a bandwidth of 6.25 MHz and a total of 16 spectral
channels. For the 330 MHz data from 2002 we used Cygnus A as a bandpass
calibrator, 3C 48 as a flux calibrator, and initial phase calibration
was done with 2321$-$163. The 2003 observations at 330 MHz used the
same bandpass and flux calibrators, and initial phases were calibrated
with 3C 48.

The data analysis procedure followed the standard spectral line
analysis for cm-wavelength data. The RFI environment at frequencies
around 330 MHz consists mainly of narrow-band interference
spikes. Flagging was initially done on the Stokes $V$ data as
astrophysical sources show very little circular polarization while RFI
is often highly circularly polarized. Additional flagging was
undertaken to remove data with large positive or negative deviations
from surrounding data. We also made use of wide-field imaging
techniques to map the primary beam area to reduce the effects of
confusing sources.

In addition to our new observations, we have also analyzed archival
VLA observations of PKS~2322$-$123 taken at 1.3 and 5 GHz with the VLA
in phased array.  This data set was taken in continuum mode and was
reduced following standard cm-wavelength techniques. Both data sets
used 3C 286 as a flux calibrator, and 2246$-$121 for phase
calibration. All radio observations are listed in
Table~\ref{tbl:radio_obs}.

\section{Radio and X-ray Interactions}

\begin{figure}[tb]
\plotone{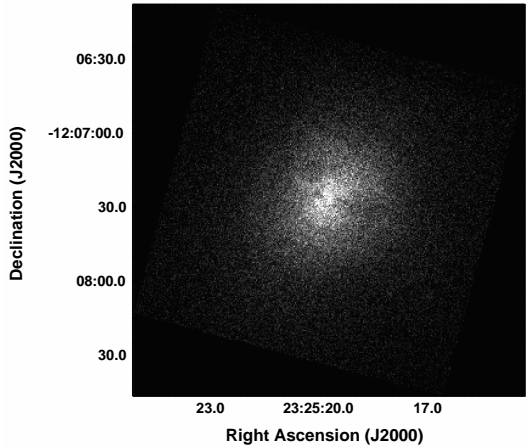}
\caption{Raw $Chandra$ image of the central $200 \times 200$ kpc
region of Abell 2597 in the $0.3-10.0$ keV energy band. No corrections
have been made for either background or exposure. Several structures
are seen in the diffuse cluster emission, including the outer
``ghost'' cavities of \citet{mcnamara01}, as well as a possible inner
cavity and an X-ray tunnel.
\label{image:raw}}
\end{figure}

\begin{figure}[tb]
\plotone{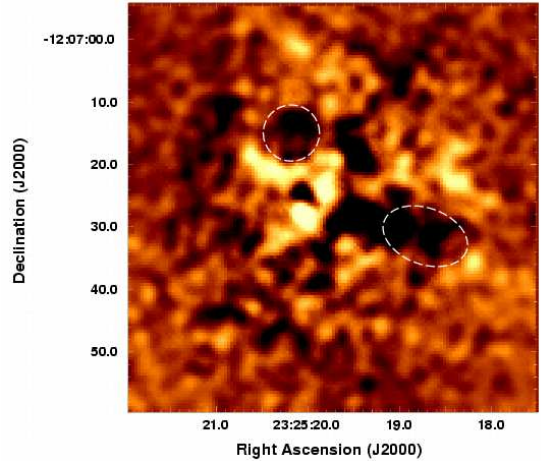}
\caption{Residual $Chandra$ image created by subtracting a smooth
elliptical model from the Gaussian smoothed ($\sigma=0\farcs98$)
image. The bright (dark) areas show regions of X-ray enhancement
(deficit) compared to the model. The two dashed regions show the
approximate size and location of the northeast and west ``ghost''
cavities reported by \citet{mcnamara01}. The western ghost cavity is
at the end of an X-ray tunnel which connects to the radio core. There
is also an inner cavity to the south of the northeast ``ghost'' cavity.
\label{image:resid_mcn}}
\end{figure}

We show the raw $Chandra$ image of the central $200 \times 200$ kpc of
Abell 2597 in Figure~\ref{image:raw}. This image shows the outer
``ghost'' cavities seen by \citet{mcnamara01} as well as additional
core structure suggestive of inner cavities.  To investigate the core
structure in more detail we have fitted a smooth elliptical model to
the data and subtracted it to produce the residual image in
Figure~\ref{image:resid_mcn}. The model fit was done using Gaussian
smoothed ($\sigma=0\farcs98$) data in the IRAF STSDAS task {\it
ellipse}. The fits allowed the ellipticity, intensity, and position
angle of the elliptical isophotes to vary within each annulus but kept
the annuli centered at the position of the compact radio core. The
residual image shows the central $100 \times 100$ kpc region of the
cluster, with a number of X-ray excess (bright) and deficit (dark)
regions visible. The dashed lines roughly outline the ``ghost''
cavities from \citet{mcnamara01}.  In our residual image, the cavity
to the west appears to be part of an X-ray tunnel running from the
radio core to a projected radius of roughly 34 kpc (22\arcsec) at a
position angle (measured from north to east) of $\sim 254$\degr.  The
\citeauthor{mcnamara01} cavity to the northeast is less striking,
although there appears to be a partial rim to the south of it.
Between the core and northeast cavity there is a compact, roughly
1\farcs5 radius, depression surrounded by regions of X-ray excess
which may represent an inner cavity.  Note that all of these features
are visible in the raw image (Figure~\ref{image:raw}), although the
more extended ones are more evident in the smoothed residual image
(Figure~\ref{image:resid_mcn}).

We estimated the significance of the western X-ray tunnel by comparing
the $Chandra$ $0.3-10.0$ keV counts in a 30\degr\ pie-wedge of an
annulus with width equal to the length of the tunnel and centered on
the cluster core, to the counts in the remainder of the annulus. This
method provides a conservative estimate that the tunnel is an
18$\sigma$ deficit over the remainder of the annulus. Using a similar
approach to determine the significance of the eastern bubble by
comparing the counts in a 30\degr\ pie wedge on the hole to the
remainder of the annulus (excluding the 30\degr\ wedge on the tunnel)
we find it represents a 7$\sigma$ deficit. 

\begin{figure}[tb]
\plotone{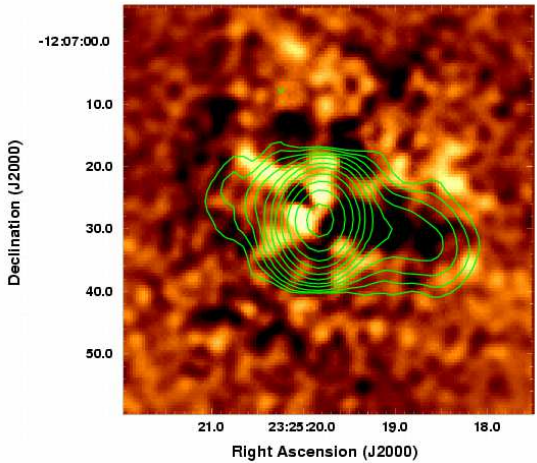}
\caption{Residual $Chandra$ image with VLA 330 MHz contours
overlaid. In addition to the compact central source, the low frequency
radio data show an extension running from the core along the X-ray
tunnel into the ``ghost'' cavity, and a possible shorter extension to
the east.  The rms level for the 330 MHz image is 1.12 mJy
beam$^{-1}$, and the restoring beam was 7\farcs5 $\times$ 5\farcs4 at
a position angle of -20\fdg6. The contours increase by factors of 2
from 4 times the rms level.
\label{image:resid+P}}
\end{figure}

In Figure~\ref{image:resid+P} we overlay our 330 MHz A configuration
contours on the X-ray residual image. Although the compact central
structure is unresolved at these low frequencies, there is a radio
extension running to the west such that the total source size is
$\sim$ 68 kpc. Figure~\ref{image:resid+P} shows a striking correlation
between the western radio extension and the X-ray tunnel.  A similar,
although less correlated, low frequency connection is seen with the 74
MHz extensions toward the ``ghost'' cavities in Perseus
\citep{perseus}.  There is also a shorter extension to the east,
although with our current data, we find no evidence of low frequency
radio emission filling the northeastern ``ghost'' cavity.  The total
flux of the radio source at 330 MHz is $8.39 \pm 0.01$ Jy.

The inner radio lobe structure of the source is seen in
Figure~\ref{image:resid+L} where we overlay our two highest resolution
radio data sets (5.0 and 1.3 GHz) on the residual
image. \citet{pollack} present the details of the spatial and spectral
properties of the inner lobes at frequencies of 5, 8, and 15 GHz.  We
note that even higher resolution VLBA observations by \citet{taylor99}
reveal symmetric inner jets at position angles of $\sim 65\degr$ and
$- 114\degr$.  The compact 5 GHz radio source has a total extent of
6\farcs7 (10 kpc) and a total flux of $364.7 \pm 0.5$ mJy. The core of
the 5 GHz radio source sits on the border of the complex X-ray
residual structure with the southern jet and lobe located at the edge
of the X-ray tunnel, and the northern lobe bounded to the north and
east by regions of excess and to the northeast by a region of X-ray
deficit.  Given the compact nature of the 5 GHz source, it may be very
difficult, even with deeper $Chandra$ observations, to image the
region around the inner radio structure accurately enough to determine
if the inner lobes are interacting with the thermal gas.  The 1.3 GHz
radio contours surround the 5 GHz source and reveal an extension
toward the east to a distance of roughly 16 kpc from the compact radio
core.  This emission appears to extend along a region of X-ray deficit
into the (compact) X-ray hole located beyond the northern 5 GHz radio
lobe.  There is a region of X-ray excess at the end of the radio
extension.  The C-shaped central 1.3 GHz emission also shows a small
excess toward the northwest, as well as weak features to the west
which may trace the western 330 MHz radio extension.

\section{Discussion}

\begin{figure}[ht]
\plotone{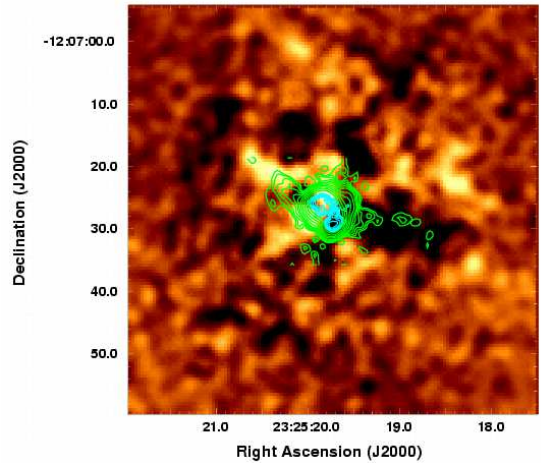}
\caption{Residual $Chandra$ image with VLA 1.3 GHz contours in green
and 5 GHz contours in cyan. Only the compact central source is seen at
5 GHz, while the 1.3 GHz data show extensions to the east, northwest,
and west.  The eastern extension is coincident with an inner X-ray
bubble. The rms level for the 5 GHz image is 55 $\mu$Jy beam$^{-1}$
and that of the 1.3 GHz data is 0.11 mJy beam$^{-1}$. The contours
increase by factors of 2 from 4 times the rms levels, although the
central contours of the 1.3 GHz data have been excluded for
clarity. The 1.3 GHz restoring beam was 2\farcs1 $\times$ 1\farcs9 at
a position angle of -17\fdg3, and the 5 GHz beam was 0\farcs5
$\times$ 0\farcs4 at a position angle of 17\fdg5.
\label{image:resid+L}}
\end{figure}

Our new radio observations of the central source in A2597 show that
the well-known compact 8 GHz source has complex radio structure at
lower frequencies.
At the highest resolution the source contains an
inverted spectrum core and symmetric radio jets extending to the
northeast and southwest \citep{taylor99}.
On larger scales, the
southwest jet appears to undergo a direction change to the south at a
distance of $\sim$ 0.8 kpc from the core before merging into the
southern inner lobe \citep{sarazin95}.
Although the northern lobe has
a bright ridge to the north, Sarazin et al.\ find no clear evidence on
larger scales of the counterjet at 8 GHz.
Similar morphology for the
compact inner structure is seen in our 5 GHz image which shows a well
resolved compact source.

Lower frequency observations at 1.3 GHz show the bright C-shaped inner
source is surrounded by extensions to the northwest, west, and east
(Figure~\ref{image:resid+L}). The eastern extension is surrounded by
bright X-ray emission and is coincident with an X-ray deficit, thus it
may represent emission from a previous outburst which has risen
buoyantly beyond the lobes of the higher frequency source. Using the
observed separation of $\sim$ 10 kpc between the eastern extension and
inner lobe, a buoyant lobe rising at 60\% of the sound
speed\footnote{\citet{ensslin02} estimate the buoyancy rise time of
radio plasma to be $v_b \propto \sqrt{r_b/r_c}\,c_s$, where $r_b$ is
the bubble radius, $r_c$ is the cluster core radius, and $c_s$ is the
sound speed. We use the 330 MHz emission to estimate $r_b$ and obtain
the core radius from \citet{pollack}.} ($c_s=350$ km s$^{-1}$) would
reach the observed (projected) location on a timescale $\tau_{buoy} >
5 \times 10^7$ yr.

Although it has been known for some time that the compact C-shaped
source in A2597 is surrounded by H$\alpha$ and [\ion{O}{2}] filaments
\citep{koekemoer}, recent $HST$ STIS FUV images by \citet{odea04}
reveal filaments of diffuse Ly$\alpha$ emission extending to radial
distances of 10\arcsec\ (15 kpc) from the radio core. A comparison of
the Ly$\alpha$ image with our 1.3 GHz image reveals that the
northeastern Ly$\alpha$ filament traces the outer edge of the
eastern 1.3 GHz radio extension (Clarke et al., in
preparation). The curved Ly$\alpha$ filament morphology and its
connection to an X-ray deficit in A2597 are reminiscent of the
H$\alpha$ filament associated with the outer radio bubbles in Perseus
which was used to place limits on the turbulence in the ICM
\citep{per_ha}.

Observations of the central source at 330 MHz reveal radio emission
extending 25\arcsec\ (38 kpc) to the west of the radio core. We also
see indications of this emission, at a very low level, in the 1.3 GHz
data. The location and spatial distribution of the 330 MHz radio
emission is well correlated with the X-ray tunnel seen as a deficit in
the $Chandra$ residual images (Figure~\ref{image:resid+P}). The
current radio data do not have sufficient resolution to determine if
the radio extension is connected to the central radio source, or if it
represents a detached buoyant radio lobe. It is interesting to note
that the symmetric VLBA jets of \citet{taylor99} extend along a
position angle that is aligned to within $\sim$ 4\degr\ of the angle
along the center of the tunnel from the radio core. 

The X-ray tunnel connected to the radio core in A2597 may represent an
intermediate case between bubbles created by an active galaxy and
those resulting from buoyant lobes.  We have used our 330 MHz radio
data to estimate the minimum energy synchrotron properties of the
western extension. We assume that the emission from the extension
comes from a uniform prolate cylinder with filling factor of unity,
and that there is equal energy in relativistic ions and electrons. The
calculations use a spectral index of $-2.7$ determined between 330 MHz
and 1.3 GHz for the tunnel, and a radio extension size of $\sim 25
\times 18$ kpc. Using a model with the magnetic field perpendicular to
the line of sight, and lower and upper cutoff frequencies of 10 MHz
and 100 GHz, we estimate minimum energy magnetic field strength of
$B_{me} = 29\ \mu$G, minimum energy non-thermal pressure of $P_{me}=5
\times 10^{-11}$ dyn cm$^{-2}$, and synchrotron lifetime of $\tau_{me}
\sim 8 \times 10^6$ yr.  Comparing this to the buoyancy rise time (at
0.6$c_s$) of $\tau_{buoy} \sim 2 \times 10^8$ yr for the bubble to
move from the projected location of the inner southern lobe to the end
of the tunnel, suggests that the electrons have been (re)accelerated
in-situ in the X-ray tunnel. Although the buoyancy rise time is likely
an underestimate due to the unknown projection effects, we note also
that the radio spectrum may flatten below 330 MHz along the western
extension, and thus the synchrotron lifetime of the particles may be
somewhat longer than calculated above. Assuming a spectral index of
$-1.0$, minimum energy estimates give a synchrotron lifetime of
$\tau_{me} \sim 5 \times 10^7$ yr. More detailed radio and X-ray
observations are required to further investigate these timescales.

The high frequency observations reveal that the
inner southern jet deflects toward the south at the location of a
bright Ly$\alpha$ blob \citep{odea04}. It is possible that the western
extension was previously fed by the radio jet prior to deflection, or
it may represent a buoyantly rising lobe from a previous
outburst. Alternatively, if the inner radio structure is porous, the
tunnel may be fed by a continuous flow of radio plasma from the inner
jets as suggested by \citet{M87} for M87.  We will undertake a
spectral study of this emission from deeper radio observations in a
future paper.

\acknowledgments

We thank Greg Taylor for helpful comments. Support for this work was
provided by the National Aeronautics and Space Administration through
{\it Chandra} awards GO2-3160X, GO3-4155X, GO3-4160X, GO4-5133X,
GO4-5149X, and GO4-5150X issued by the Chandra X-ray Observatory,
which is operated by the Smithsonian Astrophysical Observatory for and
on behalf of NASA under contract NAS8-39073. The National Radio
Astronomy Observatory is a facility of the National Science Foundation
operated under a cooperative agreement by Associated Universities,
Inc. Basic research in radio astronomy at the Naval Research
Laboratory is supported by the Office of Naval Research.

\end{document}